\documentstyle[aps,preprint,epsfig]{revtex}
\tighten
\begin{document}
\preprint{\rm FIU-NUPAR-I-\today{}}
\title {\bf Exclusive Electro-Disintegration of {\boldmath $^3$}He 
 at high {\boldmath $\boldmath Q^2$: \\ I.~Generalized Eikonal Approximation}} 
\author{M.M.~Sargsian, T.V.~Abrahamyan}
\address{Department of Physics, Florida International University, Miami, 
        Florida 33199}
\author{M.I.~Strikman}
\address{ Department of Physics, Pennsylvania State University, University Park, PA }
\author{L.L.~Frankfurt}
\address{Department of Nuclear Physics, Tel Aviv University, Tel Aviv, 
        Israel}
\maketitle
\medskip
\medskip  
\centerline{{\rm \today{}}}

\begin{abstract}
We develop a theoretical framework for calculation of high $Q^2$ exclusive electro-disintegration 
of $A=3$ systems. The main result of this work is the calculation of the final state 
interaction of the struck energetic nucleon with recoil nucleons within the generalized eikonal 
approximation (GEA) which allows us to account for the finite and relatively large  momenta of 
the bound nucleons in the nucleus. The important advantage of this approach is the possibility to 
study in a self-consistent way  the short range correlations in nuclei. This is due to the 
fact that GEA does not require a stationary condition for recoil nucleons as conventional Glauber 
approximation does. As a result GEA  treats explicitly the Fermi motion of recoil nucleons in the 
nucleus.

\end{abstract}

\section{Introduction}
\label{I}

Advances in experimental studies of high energy exclusive 
electro-disintegration reactions of few-nucleon systems\cite{LW,WP,KG,KE}  
as well as the multitude of  the planned experiments at  Jefferson Lab 
with the upgraded energies of CEBAF\cite{LC,HNM} emphasize strongly the need for 
systematic theoretical studies of these reactions. Recently, there have been 
several theoretical works addressing many  of the outstanding issues related to the 
physics  of high-energy exclusive break-up of few-nucleon systems
\cite{deuteron,SJ,Ciofi,Adam,Ryckebusch,CK,Braun1,Braun2,CKGGA,Laget2}.
 
The heightened interest in these reactions is based on expectations 
that the high resolution power  of the energetic  probe (virtual photon) and the relative 
simplicity of the target (consisting of two or three nucleons) will boost 
considerably our ability to probe the dynamics of bound 
systems at small  space-time separations and allow systematic 
studies of transition from hadronic to quark-gluon degrees of freedom 
in nuclear interactions. In many instances, these studies can take 
advantage of recent progresses in developing the realistic wave functions 
of few nucleon systems\cite{d2,Bochum}.

In this work we are interested particularly in high $Q^2$ 
($4\gtrsim Q^2\gtrsim 1$~GeV$^2$) exclusive $^3He(e,e'NN)N$ 
reactions in which one nucleon in the final state can be clearly identified
as a knocked-out nucleon which carries practically all of the  momentum of 
the virtual photon.  
We calculate the scattering amplitude of this reaction within the 
generalized eikonal approximation~(GEA)\cite{deuteron,ggraphs,treview} in which 
one starts by expressing the  scattering amplitude through the sum of the diagrams
corresponding to the $n$'th order rescattering of the knocked-out  nucleon with 
the residual nucleons in the nucleus. Then we  evaluate each diagram 
based on the effective Feynman diagram rules derived within the GEA\cite{ggraphs,treview}.
The advantage of this approach is that the number of the  diagrams 
contributing to the scattering amplitude  is defined by the finite number of $NN$ rescatterings 
that can be evaluated within the eikonal approximation. 
The manifestly covariant nature of Feynman diagrams 
allows us to preserve both the relativistic dynamics and the kinematics of the 
rescattering while identifying the low momentum nuclear part of the amplitude
with the nonrelativistic nuclear wave function. Such an approach allows us 
to account for the internal motion of residual target nucleons in the rescattering 
amplitude (for recent developments see also Ref.\cite{CKGGA}). 
Additionally, the GEA accounts for the transferred longitudinal momentum  
in the rescattering amplitude which is important for the description of the 
inelastic processes (or processes with large excitation energies) 
in which the final state is strongly energy dependent. 
All these constitute a generalization of the conventional 
eikonal approximation\cite{Glauber} in which the nucleons in the nucleus are 
considered as a stationary scatterers and only the transverse momentum is 
transferred in the reinteractions. 
These features of the GEA is crucial in describing  
electro-production reactions aimed at the study of short-range nuclear 
configurations since they are characterized by non-negligible 
values of bound nucleon momenta and excitation energies.
 The study of short-range nucleon correlations is the main goal of 
the part-II of this work. 

In the present paper we work in the  virtual nucleon framework 
in describing the dynamics of the reaction. In this framework one describes the 
reaction in the Lab frame relating  all non-nucleonic degrees of 
freedom effectively to the off-shellness of the knocked--out (virtual) nucleon in the nucleus.  
This justifies the use of only nucleonic degrees of freedom in the ground state 
wave function of the nucleus. If the probed internal momenta are 
sufficiently small,  
${p^2\over m_N^2}\ll 1$, one can use  the nonrelativistic ground state nuclear wave functions 
which are calculated based on realistic N-N interaction potentials. 
Furthermore, considering only the kinematics of quasi-elastic reactions we neglect 
the Isobar Current and Meson Exchange contributions. All these impose  
specific restrictions on the kinematics of the reaction which will be discussed 
in details in the present paper\footnote{The relativistic effects can be 
incorporated self-consistently in GEA using the light-cone formalism, see appendix in Ref.\cite{ggraphs}. 
These and studies of Isobar contributions in the reaction 
dynamics will be addressed in the subsequent papers.}.

The paper is organized as follows: In Sec.~\ref{II} the specifics 
of the considered electro-nuclear reaction and its kinematical requirements are discussed.
In Sec.\ref{III} we derive the scattering amplitude within the GEA by calculating  
the contributions from single and double rescattering of the knocked-out nucleon 
off recoil nucleons in the reaction. We calculate also the pair distortion 
effects due to the interaction of slow residual nucleons in the final state of the 
reaction. In the last part of the Sec.\ref{III} we discuss the general form of the 
differential cross section. Section \ref{IV} summarizes the results of the derivations. 
The effective Feynman diagram  rules of the GEA are given in Appendix A.

\section{Reaction and Kinematics}
\label{II}

We are considering the electro-disintegration of $^3He$ in the reaction:
\begin{equation}
e + ^3He \rightarrow e' + N_f + N_{r2}+ N_{r3}
\label{reaction}
\end{equation}
where $e$ and $e'$ are the initial and scattered electrons with four-momenta
$k_{e}$ and $k^\prime_{e}$ respectively.  The $^3$He nucleus has a four-momentum $p_{A}$.
$N_f$, $N_{r2}$ and $N_{r3}$ correspond to knocked-out and two recoil nucleons 
with four-momenta $p_f$, $p_{r2}$ and $p_{r3}$ respectively. We define also the 
four-momentum of the virtual photon $q=(q_0,{\bf q},0_\perp)\equiv k_{e}-k_{e'}$ with $Q^2=-q^2$.
Hereafter the $z$ direction is chosen parallel to $\bf q$ and the scattering plane is  
the plane of the $\bf q$ and ${\bf k_{e'}}$ vectors.

We will investigate  the reaction of Eq.(\ref{reaction}) 
in the kinematic  region defined as follows:
\begin{eqnarray}
\mbox{(a)} \ 4 \ge Q^2\ge 1 \mbox{GeV}^2; \ \ \ \
\mbox{(b)} \ {\bf p_{f}}\approx {\bf q}; \ \ \ \ 
\mbox{(c)} \ |{\bf p_m}|,|{\bf p_{r2}}|, |{\bf p_{r3}}| \le 400-500 \mbox{MeV/c}, \ \ \ \
\label{kin}  
\end{eqnarray}
where one defines a missing momentum ${\bf p_m} = {\bf p_f} - {\bf  q}$.
The lower limit of Eq.(\ref{kin})(a) is what  
provides a high-momentum transfer in the electro-disintegration. This condition together with  
Eqs.(\ref{kin})(b) and (c)  allows us to identify $N_f$ as a  knocked-out nucleon, 
while $N_{r2}$ and $N_{r3}$ could be considered as recoil nucleons which do not 
interact directly with the virtual photon.  The upper limit of Eq.(\ref{kin})(a) comes
from the condition that the color coherence effects  are small and the 
produced hadronic state represents a single state (i.e. nucleon) rather than   
a  superposition of different hadronic states  in the form of the wave packet
(see e.g. Ref.\cite{FMS94}). 

Additionally, the  condition of Eq.(\ref{kin})(c) allows us to consider  
the nucleons as the basic degrees of freedom in describing the interacting 
nuclear system. From the technical point of view, this means that in the set of 
noncovariant diagrams comprising the covariant scattering amplitude, one  can neglect 
the noncovariant diagrams containing non-nucleonic degrees of 
freedom (e.g. negative energy projections of the bound nucleon spinors contributing 
to the vacuum fluctuations in the scattering amplitude). Within this approximation 
one can reduce the nuclear vertices to the nonrelativistic nuclear wave functions of nuclei 
(see e.g. Eq.(\ref{wf})). Note that on several occasions in Ref.\cite{SASF2} we will extend 
our calculations to the region of missing and recoil momenta $\ge 500$~MeV/c. We justify such 
an extension by the expectation that the onset of the relativistic 
effects in the nuclear wave function should happen rather smoothly. 
However, in all these cases our results should be considered as qualitative.

\begin{figure}[ht]
\centerline{\epsfig{height=4cm,width=9cm,file=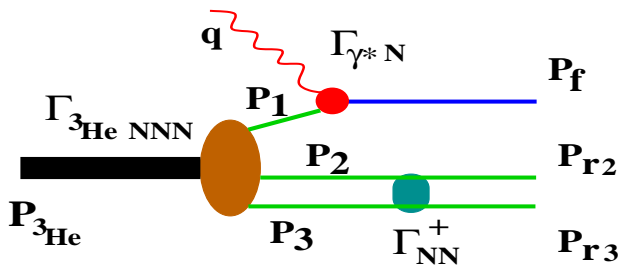}}
\caption{(Color online) Impulse Approximation contribution to the scattering amplitude of 
         $^3He(e,e'N_fN_{r2})N_{r3}$ reaction.}
\label{IAfig}
\end{figure}

\section{Scattering Amplitude}
\label{III}

Within the one photon exchange approximation, the amplitude, $M_{fi}$
of reaction (\ref{reaction}) can be written as follows
(see e.g. \cite{Feynman}):
\begin{equation}
M_{fi}=-4\pi \alpha \frac{1}{q^2} j^e_{\mu}\cdot
\langle f| J^{\mu}_{A}(Q^2)| i \rangle \ .
\label{ampl}
\end{equation}
Here  $\alpha$ is the fine structure constant;
$j^e_\mu=\bar{u}(k_{e}^{\prime})\gamma_{\mu}u(k_{e})$ is the electromagnetic
current of the electron and  $J^{\mu}_{A}(Q^2)$ is the operator of the nuclear electromagnetic  
current. The initial state $|i\rangle$, which enters Eq.\ (\ref{ampl}), is the totally
antisymmetric state of $^3$He. The final state  $|f\rangle$ also has to be antisymmetric. 
However because of the kinematical constraints of Eq.(\ref{kin}) one can neglect the 
antisymmetrization  between the outgoing fast and slow recoil nucleons. 
Such an approximation is justified due to the fact that the diagrams in which an 
energetic photon will produce the slow hadrons are strongly suppressed.

Here we need to calculate the electromagnetic transition amplitude 
$A^{\mu}$, defined as
\begin{equation}
A^\mu \equiv \langle f| J^{\mu}_{A}(Q^2)| i \rangle \ .
\label{Amu}
\end{equation}

We will perform the calculation  within the generalized eikonal approximation~(GEA), 
\cite{ggraphs,treview}.  In this approach, the interaction  of the fast, 
knocked-out nucleon with slow spectator nucleons  is calculated 
based on the  eikonal approximation for  the corresponding  covariant 
diagrams. The reduction theorem\cite{treview} derived for this approximation 
allows us to reduce an infinite sum of rescattering diagrams to a finite set of 
covariant Feynman diagrams. In these diagrams,  
soft NN reinteractions are described through phenomenological NN interaction vertices 
which can be parameterized using experimental data on the  small angle NN scattering amplitudes. 
In its final form this approximation can be formulated through a  set of effective Feynman 
diagram rules (summarized in Appendix A) for calculating the scattering amplitude 
of the $e+A\rightarrow e' + N + (A-1)'$ reaction in the given order of the rescattering of 
fast knocked-out nucleon off spectator nucleons (for review on GEA, see Ref.\cite{treview}). 
Based on the kinematic constraints of  Eq.(\ref{kin})(c) we 
will neglect non-nucleonic  degrees of freedom in the ground state wave function 
of $^3He$. This allows us, in the limit of $p_m^2/m^2\ll 1$, to reduce the covariant nuclear 
vertices to the nonrelativistic wave function of initial  and final  nuclear states with nucleonic 
constituents only.  Note that we still account for effects of the order of magnitude $p_m/m$. 
One such term is the flux factor which is proportional to $1-p^z_m/m$ and which 
should be taken into account to preserve the baryon number conservation in the reaction 
(see e.g. Ref.\cite{FS76}). In addition to the flux factor effects, in the present 
approach, the initial  off-shellness of the struck nucleon  
renders ambiguity in choosing the proper form of the electromagnetic current of the $eN$ 
interaction.  This problem is addressed usually by  imposing an electromagnetic current conservation 
relation that allows us to express  the off-shell component through the on-shell component of the 
electromagnetic current (see e.g.\cite{deFor}). Note, however, that the ambiguity due to the 
off-shellness  is proportional to ${p_m^2\over Q^2}$, and for the kinematic range of 
Eq.(\ref{kin}), it is a small correction the discussion of which is out of scope of the 
present paper.

\subsection{Impulse Approximation}
\label{IAsec}

The contribution to the electromagnetic transition amplitude $A^{\mu}$,
in which the knocked out nucleon does not interact with other nucleons, corresponds to the 
impulse approximation (IA). In this case the wave function of 
the knocked  out nucleon is a plane wave. 

The IA term of  the scattering amplitude, $A_0^{\mu}$, is described by the 
Feynman diagram of  Fig.\ref{IAfig}. Using the diagrammatic rules summarized in Appendix~A
and identifying  knocked-out and two recoil nucleons in the initial  state by 
indexes $1$, $2$ and $3$ respectively, one obtains:

\begin{eqnarray}
A_0^\mu &=& -\int {d^4p_2\over i (2\pi)^4} \bar u(p_{r2})\bar
u(p_{r3})\bar u(p_f)\cdot \Gamma_{NN}^{+}(p_2,p_3)\cdot \Gamma^{\mu}_{\gamma^*N }
\cdot {\hat p_3+m\over p_3^2-m^2 + i\varepsilon} \times \nonumber  \\ &\times& 
{\hat p_2+m\over p_2^2-m^2 + i\varepsilon} \cdot {\hat p_1+m\over p_1^2-m^2 + 
i\varepsilon} 
\cdot \Gamma_{ ^3 \mathrm{He} NNN}(p_1,p_2,p_3) \: \chi^{A},
\label{IAeq1}
\end{eqnarray}
where ${\bf p_1} = {\bf p_m}\equiv {\bf p_f}- {\bf q}$ and $p_3 =  p_{A}-p_1-p_2$.
For the kinematic range of  Eq.(\ref{kin})(c) one can integrate over $dp^0_{2}$  estimating 
it through the residue at the positive energy pole of the propagator of the nucleon ``2''. 
This   corresponds to a  positive energy projection of the virtual nucleon state.  
Such integration  effectively corresponds  to the replacement:

\begin{equation}
\int {dp^0_2\over p_2^2-m^2+i\varepsilon} = -{i2\pi\over 2 E_2}\approx -{i2\pi\over 2 m}.
\label{residue}
\end{equation}
The condition that the internal momenta of the nucleons remain small (${\bf p^2_{m,2,3}}\ll m^2$) 
also allows one to use the closure relation for on-mass shell nucleons to express the numerator of 
the bound nucleon propagator as follows:
\begin{equation}
\hat p + m = \sum\limits_{s}u(p,s)\bar u(p,s).
\label{closure}
\end{equation}

Using Eqs.(\ref{residue}) and (\ref{closure}) in Eq.(\ref{IAeq1}) one obtains:
\begin{eqnarray}
A^\mu_{0}&=& \sum_{s_1s_2s_3} \int {d^3p_2\over 2m (2\pi)^3} \: 
{\bar{u}(p_{r2},s_{r2})\bar{u}(p_{r3},s_{r3})\Gamma_{NN}^{+}(p_2,p_3)u(p_3,s_3)u(p_2,s_2)
\over p_3^2-m^2 + i\varepsilon} \times \nonumber \\ 
& \times&  \bar{u}(p_f,s_f)\Gamma^{\mu}_{\gamma^*N}u(p_1,s_1) 
{\bar {u}(p_1,s_1)\bar {u}(p_2,s_2)\bar {u}(p_3,s_3) 
\Gamma_{^3 \mathrm{He}NNN}(p_1,p_2,p_3) \: \chi^{A} \over p_1^2-m^2 + 
i\varepsilon}.
\label{IAeq2}
\end{eqnarray}
Using Eq.(\ref{wf}) and introducing the electromagnetic nucleon current,
\begin{eqnarray}
j^\mu(p_f,s_f;p_m,s_1) &=& \bar u(p_f,s_f)\Gamma^{\mu}_{\gamma^*N}u(p_m,s_1),
\label{j_cur}
\end{eqnarray}
for $A_0^\mu$ one arrives at:
\begin{eqnarray}
A^\mu_{0}&=& \sqrt{2E_{r2}2E_{r3}}(2 \pi)^3 \sum_{s_1,s_2,s_3,t_2,t_3}\int d^3 p_2 
\Psi_{NN}^{\dagger p_{r2},s_{r2},t_{r2};p_{r3},s_{r3},t_{r3}}(p_2,s_2,t_2;p_3,s_3,t_3)\nonumber \\ 
& & \times j_{t1}^{\mu}(p_m+q,s_f;p_m,s_1) 
\Psi^{s_A}_{A}(p_{m},s_1,t_1;p_{2},s_2,t_2;p_{3},s_3,t_3),
\label{IAeq3}
\end{eqnarray}
where $s_A,s_1,s_2,s_3,s_f,s_{r2},s_{r3}$ describe the spin projections of the $^3He$-nucleus, 
the initial nucleons and the final nucleons respectively. We 
represent the  isospin projections of nucleons by $t_1,t_2,t_3,t_{f},t_{r2},t_{r3}$ 
and use these indexes to identify the protons and neutrons.
In the above equation,  $\Psi^{s_A}_{A}$ is the ground state wave 
function of the $^3He$ nucleus with polarization  vector ${\bf s_A}$ and $\Psi_{NN}$ represents the 
bound or continuum  $NN$ wave function.  One can simplify further Eq.(\ref{IAeq3}) 
using the fact that $\Psi_{NN}$ is a function only of the relative three-momenta of spectator 
nucleons and the spins.  This allows us to replace the $d^3p_2$ integration by $d^3 p_{23}$
which yields 
\footnote{To do this one can introduce $d^3p_3\delta^3(p_{r2}+p_{r3}-p_{2}-p_{3})$ in 
Eq.(\ref{IAeq3}), then replace $d^3p_{2}d^3p_{3}$ by $d^3p_{23}d^3P_{cm23}$, 
with ${\bf p_{23}}= {\bf {p_2-p_3\over 2}}$ and ${\bf P_{cm23}}={\bf p_2+p_3}$  
and integrate out the $\delta$ function through $d^3P_{cm23}$.}:
\begin{eqnarray}
A^\mu_{0}&=& \sqrt{2E_{r2}2E_{r3}}(2 \pi)^3 \sum_{s_1,s_2,s_3,t_2,t_3}\int d^3 p_{23} 
\Psi_{NN}^{\dagger p_{r23},s_{r2},t_{r2};s_{r3},t_{r3}}(p_{23},s_2,t_2;s_3,t_3)\nonumber \\
& & \times  j_{t1}^{\mu}(p_m+q,s_f;p_m,s_1) 
\Psi^{s_A}_{A}(p_{m},s_1,t_1;-{p_m\over 2}+p_{23},s_2,t_2;-{p_m\over 2}-p_{23},s_3,t_3).
\nonumber \\
\label{IAfin}
\end{eqnarray}
For the case of the reaction of Eq.(\ref{reaction}), $\Psi_{NN}$ 
is a continuum $NN$ wave function which can be represented through 
the Lippmann-Schwinger equation as follows:
\begin{eqnarray}
\Psi_{NN}^{\dagger p_{r23,s_{r2},t_{r2};s_{r3},t_{r3}}}(p_{23},s_2,t_2;s_3,t_3)
 &=&\delta^3({\bf p_{23}}-{\bf p_{r23}})+ \nonumber \\
& & {1\over 2\pi^2}{\langle s_{r2},t_{r2};s_{r3},t_{r3}\mid f_{NN}^{\mathrm{off \: shell}}
({\bf p}_{r23},{\bf p}_{23}) 
\mid s_{2},t_{2};s_{3},t_{3}\rangle
\over {\bf p_{23}}^2-{\bf p_{r23}}^2 - i \varepsilon },\nonumber \\
\label{ypages}
\end{eqnarray}
where ${\bf p}_{23}={{\bf p}_2-{\bf p}_3\over 2}$, 
${\bf p_{r23}}={{\bf p _{r2}}-{\bf p_{r3}}\over 2}$ 
and $f_{NN}^{\mathrm{off \: shell}}$ is a half-off-shell  
non-relativistic amplitude of NN scattering~(see e.g. Ref.\cite{BJ}). 
Terms at the right-hand side of 
Eq.(\ref{ypages}) characterize two distinctive dynamics of production of the recoil NN state. 
If only the first term of Eq.(\ref{ypages}) is kept  in Eq.(\ref{IAfin}), 
this will correspond to the approximation in which all three final 
nucleons propagate as plane waves. Hereafter we will refer to this 
approximation as a plane wave impulse approximation~(PWIA). The second 
term in Eq.(\ref{ypages}) describes a reinteraction between the pair of 
the slow nucleons which distorts the plane wave of the outgoing recoil 
nucleons. Following Ref.\cite{Laget} we will refer to  this term  
as a pair distortion contribution.

\subsection{Single rescattering amplitude}
\label{RS1sec}

The diagrams in Fig.~\ref{RS1fig}  describe the process in which  the knocked-out (fast) 
nucleon rescatters  off one of the spectator nucleons. Using the diagrammatic rules from 
Appendix A, for the  amplitude corresponding to the diagram of Fig.~\ref{RS1fig}(a)
one obtains:
\begin{eqnarray}  
A_{1\mathrm{a}}^{\mu}&=&-\int {d^4 p_2 \over i (2 \pi)^4}{d^4 p_3 \over i (2\pi)^4} 
\bar{u}(p_{r3})\bar{u}(p_{r2})\bar{u}(p_{f}){\Gamma_{NN}^{+}(p'_2,p_3) (\hat p'_2+m) 
\over  p_{2}^{\prime 2}-m^2 + i\varepsilon} \times \nonumber \\
 &\times& {F_{NN}^{\mathrm{a}}(p'_2-p_2) (\hat{p}_1+\hat{q}+m) \over  
(p_1+q)^2-m^2 + i\varepsilon}\cdot \Gamma^{\mu}_{\gamma^*N } \cdot {\hat p_3 +m 
\over p_3^2-m^2+i\varepsilon} \times  \nonumber \\
&\times& {\hat p_2 +m \over p_2^2-m^2+i\varepsilon}\cdot {\hat p_1 +m 
\over p_1^2-m^2+i\varepsilon} \cdot \Gamma_{ ^3 \mathrm{He} NNN}(p_1,p_2,p_3) \: 
\chi^{A}.
\label{FSI1a}
\end{eqnarray}
Using the same arguments as in the previous section we evaluate $d^0p_{2}$ and $d^0p_{3}$
integrals  through the residues at positive energy poles in the recoil nucleon propagators. 
This yields a replacement of $\int d^0_{2,3}{1\over p^2_{2,3}-m^2+i\varepsilon} 
\approx -i{2\pi\over 2m}$ and 
reduces the covariant Feynman diagram of Eq.(\ref{FSI1a}) to the time-ordered 
noncovariant diagram in which nucleon ``1'' is first struck by a virtual photon 
and then  rescatters off the spectator nucleon, ``2''. The rescattered nucleon ``2" 
subsequently combines  with the nucleon ``3'' into  the $NN$ continuum (or bound) state.
Now, we can use Eq.(\ref{closure}) for the intermediate nucleons.  
Furthermore, relating the nuclear vertex functions to the nuclear wave functions according to  
Eq.(\ref{wf}) and the $NN$ rescattering vertex functions to the $NN$ scattering amplitude 
according to Eq.(\ref{NN}), one obtains:

\begin{figure}[ht]
\centerline{\epsfig{height=7cm,width=9cm,file=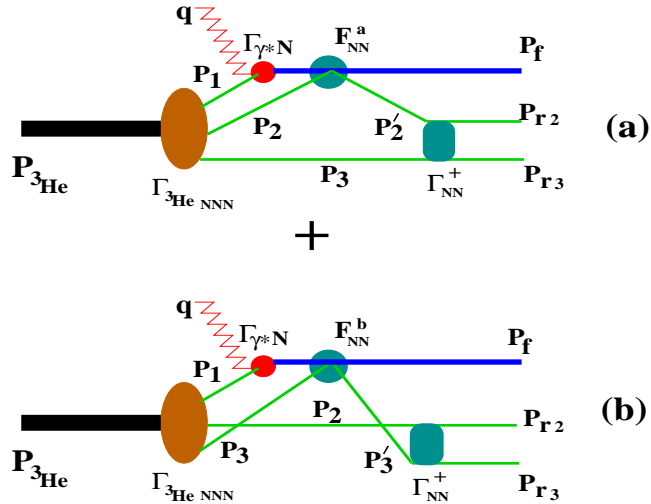}}
\caption{(Color online) Single rescattering contribution to the scattering amplitude 
         of $^3He(e,e'N_fN_{r2})N_{r3}$ reaction.}
\label{RS1fig}
\end{figure}

\begin{eqnarray}
A_{1\mathrm{a}}^{\mu}&=& -F\sum\limits_{s_{1'},s_{2'},s_{1},s_{2},s_{3}}\sum
\limits_{t_1,t_{2'},t_2,t_3}
{1 \over 2m}\int {d^3 p_2 \over (2 \pi)^3}d^3 p_3 
\Psi_{NN}^{\dagger p_{r2},s_{r2},t_{r2}; p_{r3},s_{r3},t_{r3}}(p'_2,s_{2'},t_{2'};p_3,s_3,t_3) 
\times \nonumber \\ 
&\times& {\sqrt{s^{NN}_2(s^{NN}_2-4m^2)}
f_{NN}(p'_{2},s_{2'},t_{2'},p_f,s_{f},t_f;| p_{2},s_2,t_2,p_1+q,s_{1'},t_{1})
\over(p_1+q)^2-m^2+i\varepsilon}\nonumber  \\
&\times & j_{t1}^{\mu}(p_1+q,s_{1'} ;p_1,s_1) \cdot 
\Psi^{s_A}_{A}(p_{1},s_1,t_1;p_{2},s_2,t_2;p_{3},s_3,t_3),
\label{FSI1b}
\end{eqnarray}
where $F= \sqrt{2E_{r2}2E_{r3}}(2 \pi)^3$, $p_1 = p_A-p_2-p_3$, $p'_2 = p_{r2}+p_{r3}-p_3$ 
and $s^{NN}_2 = (p_1+q+p_2)^2$.

Now we analyze the propagator of the knocked-out nucleon:
\begin{equation} 
(q+p_1)^2 - m^2 + i\epsilon = 
2q\left[{2mq_0-Q^2\over 2q} - p_{1z} + i\epsilon\right] \ .
\label{propagator1}
\end{equation}
The factor ${2mq_0-Q^2\over 2q}$ is fixed by external (measurable) 
kinematical variables such us 
$Q^2$, $q_0$, ${\bf p_f}$ and ${\bf p_{r2}}$. Using the condition of the quasi-elastic 
scattering for reaction (\ref{reaction}),
$(q+p_A-(p_{r2}+p_{r3}))^2=m^2$, one obtains:
\begin{equation}
{2mq_0-Q^2\over 2q} = p_{mz} + {q_0\over q}(T_{r2}+T_{r3}+|\epsilon_A|)
+ {m^2-\tilde m^2\over 2 q}\approx p_{mz} + \Delta^0 \ ,
\label{kinem}
\end{equation} 
where $p_{mz} = p_{fz}-q$. $T_{r2}$ and $T_{r3}$ are the kinetic 
energies of  recoil nucleons, $|\epsilon_A|$ is the modulus of the binding energy 
of the target  and $\tilde m^2 = [p_A-(p_{r2}+p_{r3})]^2$. 
In the last step in Eq.(\ref{kinem}) we neglected the  
$(m^2-\tilde m^2)/(2q)$ term, since for the 
fixed decay kinematics it vanishes with an increase of $q$.
We have  also denoted
\begin{equation}
\Delta^0= {q_0\over q} (T_{r2}+T_{r3}+|\epsilon_A|),
\label{Delta}
\end{equation}
which defines the effective longitudinal momentum transfered in the $NN$ rescattering.
Note that $\Delta^0$, which is absent in the conventional eikonal approximation, is 
important at large values of recoil nucleon energies $T_{r2}$,$T_{r2}\sim m$. As it will 
be shown in part-II of this paper\cite{SASF2} the kinematics with the large values of recoil nucleon energies are most 
relevant for accessing the short-range correlations in nuclei.
Substituting Eqs. (\ref{propagator1}) and (\ref{kinem}) 
into Eq.(\ref{FSI1b}) for the scattering amplitude described by 
Fig.\ref{RS1fig}, one obtains:
\begin{eqnarray}
A_{1a}^{\mu}&=& -{F\over 2}\sum\limits_{s_{1'},s_{2'},s_{1},s_{2},s_{3}}\sum
\limits_{t_1,t_{2'},t_2,t_3}
\int {d^3 p_2 \over (2 \pi)^3} d^3 p_3 
\Psi_{NN}^{\dagger p_{r2},s_{r2},t_{r2}; p_{r3},s_{r3},t_{r3}}(p'_2,s_{2'},t_{2'};p_3,s_3,t_3) 
\nonumber \\ 
&\times& {\sqrt{s^{NN}_2(s^{NN}_2-4m^2}) \over 2qm}
{f_{NN}(p'_{2},s_{2'},t_{2'},p_f,s_{f},t_f;| p_{2},s_2,t_2,p_1+q,s_{1'},t_1)
\over p_{mz}+\Delta^0-p_{1z}+i\varepsilon}\nonumber \\
&\times & j_{t_1}^{\mu}(p_1+q,s_{1'};p_1,s_1) \cdot 
\Psi^{s_A}_{A}(p_{1},s_1,t_1;p_{2},s_2,t_2;p_{3},s_3,t_3).
\label{FSI1c}
\end{eqnarray}
Two important features of soft $NN$ scattering allow us to 
simplify Eq.(\ref{FSI1c}). One is that in the high-energy regime the soft, low $t$, $NN$ 
amplitude, which dominates in Eq.(\ref{FSI1c}) conserves the helicities of nucleons. 
Starting at $Q^2\ge 1$~GeV$^2$ the approximation of helicity conservation is accurate on the 
level of 2-3\%\cite{Vorobev} (for recent analysis see Ref.\cite{SAID}). It is even smaller factor 
for the observables of reaction (\ref{reaction}) in which no polarization degrees of freedom are fixed.  
The second is that the soft amplitude is a function of the transverse 
component of the transfered momentum, $(p'_2-p_2)_\perp$ only.
Since  $\hat z || {\bf q}$, 
the helicity conserving argument implies the conservation of the 
polarization projections  of the interacting nucleons in the $z$ direction. 
These simplifications yield the equation,
\begin{equation}
f_{NN}(p'_{2},s_{2'},t_{2'},p_f,s_{f},t_f;| p_{2},s_2,t_2,p_1+q,s_{1'},t_1) = 
f^{t_{2'},t_f|t_2,t_1}_{NN}(p'_{2\perp}-p_{2\perp})\delta^{s_{f},s_{1'}}\delta^{s_{2'},s_{2}}.
\label{soft}
\end{equation}
Using this relation and defining the  transfered momentum in $f_{NN}$ as 
$k\equiv p'_{2}-p_{2}=p_{1}-p_{m}$ we can rewrite  Eq.(\ref{FSI1c}) as follows:
\begin{eqnarray}
A_{1\mathrm{a}}^{\mu}&=& -{F\over 2}\sum\limits_{s_{1},s_{2},s_{3}}\sum\limits_{t_1,t_{2'},t_2,t_3}
\int {d^3 k \over (2 \pi)^3} d^3 p_3 
\Psi_{NN}^{\dagger p_{r2},s_{r2},t_{r2}; p_{r3},s_{r3},t_{r3}}(p'_2,s_{2},t_{2'};p_3,s_3,t_3) \times
\nonumber \\ 
&\times& {\chi_1(s^{NN}_{2})f^{t_{2'},t_f|t_2,t_1}_{NN}(k_{\perp})
\over \Delta^0-k_z+i\varepsilon}
\cdot j_{t_1}^{\mu}(p_1+q,s_{f} ;p_1,s_1) \cdot 
\Psi^{s_A}_{A}(p_{m}+k,s_1,t_1;p_{2},s_2,t_2;p_{3},s_3,t_3),
\nonumber \\ 
\label{FSI1d}
\end{eqnarray}
where $\chi_1(s^{NN}_{2}) = {\sqrt{s^{NN}_{2}(s^{NN}_{2}-4m^2}) \over 2qm}$.

The contribution of the second diagram in Fig.\ref{RS1fig} can be calculated by 
interchanging the momenta of ``2'' and ``3'' nucleons in Eq.(\ref{FSI1d}). 
Doing this and changing the integration variables $d^3p_3$ to $d^3p_{23}$ 
(similar to Sec.\ref{IAsec})  for complete single rescattering amplitude one  obtains:
\begin{eqnarray}
A^\mu_1 & = & A_{1\mathrm{a}}^{\mu} + A_{1\mathrm{b}}^{\mu} 
\nonumber \\
& = & -{F\over 2}\sum\limits_{s_{1},s_{2},s_{3}}\sum\limits_{t_{2'}t_{3'}}
\sum\limits_{t_1,t_{2}t_{3}}
\int {d^3 k d^3 p_{23}  \over (2 \pi)^3} 
\Psi_{NN}^{\dagger p_{r23},s_{r2},t_{r2};s_{r3},t_{r3}}(p_{23},s_{2},t_{2'};s_3,t_{3'}) 
\cdot j_{t_1}^{\mu}(p_1+q,s_{f} ;p_1,s_1) 
\nonumber \\
& & \times\left[{\chi_1(s^{NN}_2)f^{t_f,t_{2'}|t_1,t_2}_{NN}(k_{\perp})\delta^{t_{3'},t_3}\over 
\Delta^0-k_z+i\varepsilon} 
\Psi^{s_A}_{A}(p_{m}+k,s_1,t_1;-{p_m\over 2} + p_{23}-k,s_2,t_2;
 -{p_m\over 2} - p_{23};s_3,t_{3}) \right.  \nonumber \\ 
& & \left. \ \  + {\chi_1(s^{NN}_3)f^{t_f,t_{3'}|t_1,t_3}_{NN}(k_{\perp})\delta^{t_{2'},t_2}\over 
\Delta^0-k_z+i\varepsilon} 
\Psi^{s_A}_{A}(p_{m}+k,s_1,t_1;-{p_m\over 2} + 
p_{23},s_2,t_{2};-{p_m\over 2} - p_{23}-k;s_3,t_3) \right].
\nonumber \\
\label{FSI1fin}
\end{eqnarray}

\begin{figure}[ht]
\centerline{\epsfig{height=7cm,width=9cm,file=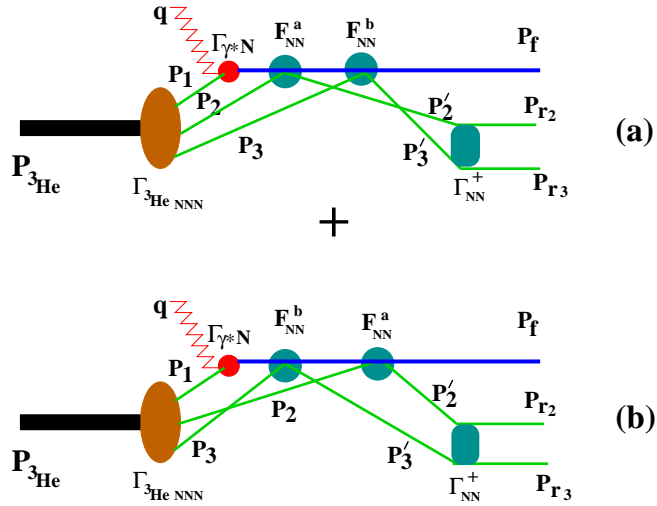}}
\caption{(Color online) Double rescattering contribution to the scattering amplitude of 
$^3He(e,e'N_fN_{r2})N_{r3}$ reaction.}
\label{FSI2fig}
\end{figure}

\subsection{Double rescattering amplitude}
\label{FSI2sec}

Next we discuss the double rescattering contribution, in which knocked-out nucleon 
rescatters off both spectator nucleons in the nucleus, Fig.\ref{FSI2fig}. 
Using the diagram rules summarized in Appendix A one obtains:
\begin{eqnarray}  
\!\!A_{2\mathrm{a}}^{\mu}&=&-\int{d^4 p'_3 \over i (2\pi)^4} 
{d^4 p_2 \over i (2 \pi)^4}{d^4 p_3 \over i (2\pi)^4} 
\bar{u}(p_{r3})\bar{u}(p_{r2})\bar{u}(p_{f}) 
{\Gamma_{NN}^{+}(p'_2,p'_3) (\hat p'_2+m)(\hat p'_3+m) \over  
(p_{2}^{\prime 2}-m^2 + i\varepsilon)(p_{3}^{\prime 2}-m^2 + i\varepsilon)} \times  \nonumber \\
&\times&  {F_{NN}^{\mathrm{b}}(p'_3-p_3) (\hat{p}_1+\hat{q}+\hat{p_2}-\hat{p'_2}+m) 
\over  (p_1+q+p_2-p'_2)^2-m^2 + i\varepsilon}\times \nonumber \\ 
&\times& {F_{NN}^{\mathrm{a}}(p'_2-p_2) (\hat{p}_1+\hat{q}+m) 
\over  (p_1+q)^2-m^2 + i\varepsilon} \cdot \Gamma^{\mu}_{\gamma^*N } 
\cdot {\hat p_3 +m \over p_3^2-m^2+i\varepsilon} \times  \nonumber \\
&\times& {\hat p_2 +m \over p_2^2-m^2+i\varepsilon}\cdot 
{\hat p_1 +m \over p_1^2-m^2+i\varepsilon} \cdot 
\Gamma_{ ^3 \mathrm{He} NNN}(p_1,p_2,p_3) \: \chi^{A},
\label{FSI2a}
\end{eqnarray}
where $p_2$ and $p_3$ are the momenta of the spectator nucleons
before rescattering; $p_1 = p_A - p_2 - p_3$.

Using the  same approximation we used for the IA and single rescattering 
amplitudes we estimate the integrals over  $d^0p_{3',3,2}$  through the positive 
energy poles of the  nucleon propagators  with momenta $p'_{3}$, $p_{3}$ and $p_2$ 
respectively. These integrations result in the replacement of 
$\int {d^0p_j\over 2\pi i (p^2_j-m^2+i\epsilon)} \rightarrow -{1\over 2E_j}\approx -{1\over 2m}$, 
($j=2,3,3'$). 

Applying the closure condition of  Eq.(\ref{closure}) and using the 
reduction (Eq.(\ref{wf}))  of nuclear vertices to the nonrelativistic nuclear wave functions 
(both in the ground state and in the continuum) as  well as applying the relations of   
Eqs.(\ref{j_cur},\ref{NN}), for $A^\mu_{2a}$ one obtains:
\begin{eqnarray}
A_{2a}^{\mu}&=& {F \over (2m)^2}\sum\limits_{s_{1},s_{2},s_{3}}
\sum\limits_{t_1,t_2,t_3,t_{1'},t_{2'},t_{3'}}
\int {d^3 p'_3 \over (2\pi)^3}{d^3 p_2 \over (2 \pi)^3} {d^3 p_3} 
\Psi_{NN}^{\dagger p_{r2},s_{r2},t_{r2};p_{r3},s_{r3},t_{r3}}(p'_2,s_{2},t_{2'};p'_3,s_{3},t_{3'})
\times \nonumber \\ 
&\times& {\sqrt{s^{NN}_{b3}(s^{NN}_{b3}-4m^2)} f^{t_{3'},t_f | t_3, t_{1'}}_{NN}(p'_{3\perp}-p_{3\perp})
\over (p_1+q+p_2-p'_2)^2-m^2+i\varepsilon} 
 {\sqrt{s^{NN}_{a2}(s^{NN}_{a2}-4m^2)} f^{t_{2'},t_{1'}|t_2,t_1}_{NN}(p'_{2\perp}-p_{2\perp})
\over (p_1+q)^2-m^2+i\varepsilon}  \nonumber \\ 
&\times& j_{t1}^{\mu}(p_1+q,s_f;p_1,s_1) 
\cdot \Psi^{s_A}_{A}(p_{1},s_1,t_1;p_{2},s_2,t_2;p_{3},s_3,t_3),
\label{FSI2b}
\end{eqnarray}
where $s^{NN}_{b3}$ and $s^{NN}_{a2}$ are total invariant energies of nucleons 
coupling at the  vertices $F_{NN}^{\mathrm{b}}$ and $F_{NN}^{\mathrm{a}}$ respectively.

Let us now consider the denominators of the knocked-out nucleon in the 
intermediate states. For $(p_1+q)^2-m^2+i\varepsilon$, similar to the case of 
single rescattering, one obtains
\begin{equation}
(p_1+q)^2-m^2 = 2q(\Delta^0+p_{mz}-p_{1z} + i\epsilon) \ ,
\label{Delta1}
\end{equation}
where $\Delta^0$ is defined according to Eq.(\ref{Delta}).

For the denominator, $(p_1+q+p_2-p'_2)^2-m^2+i\varepsilon$ in Eq.(\ref{FSI2b}) using 
energy-momentum conservation in the reaction (\ref{reaction}) we obtain,
\begin{eqnarray}
(p_1+q+p_2-p'_2)^2-m^2+i\epsilon & = & (p_{f} + p'_3-p_3)^2 -m^2 + i\epsilon  \nonumber \\ 
& & =   2p_{fz}\left[{E_f\over p_{fz}}(E'_3-E_3)-(p'_{3z}-p_{3z}-
{p_{f\perp}\over p_{fz}} (p'_{3\perp}-p_{3\perp})+i\epsilon\right]\nonumber \\
& & = 2p_{fz}\left[\Delta_{3} - (p'_{3z}-p_{3z}) + i\epsilon\right]. 
\label{Delta3}
\end{eqnarray}
where $\Delta_{3} = {E_f\over p_{fz}}(E'_3-E_3)- {p_{f\perp}\over p_{fz}} (p'_{3\perp}-p_{3\perp})$.
Substituting Eqs.(\ref{Delta1},\ref{Delta3})   into Eq.(\ref{FSI2a})
for $A^\mu_{2a}$ one obtains:
\begin{eqnarray}
A_{2a}^{\mu}&=& {F \over 4}\sum\limits_{s_{1},s_{2},s_{3}}
\sum\limits_{t_1,t_2,t_3,t_{1'},t_{2'},t_{3'}}
\int {d^3 p'_3 \over (2\pi)^3}{d^3 p_2 \over (2 \pi)^3} {d^3 p_3} 
\Psi_{NN}^{\dagger p_{r2},s_{r2},t_{r2};p_{r3},s_{r3},t_{r3}}(p'_2,s_{2},t_{2'};p'_3,s_{3},t_{3'})
\times \nonumber \\ 
&\times& {\chi_2(s^{NN}_{b3})f^{t_{3'},t_f | t_3, t_{1'}}_{NN}(p'_{3\perp}-p_{3\perp})
\over \Delta_3 + p'_{3z}-p_{3z} + i\varepsilon} 
{\chi_1(s^{NN}_{a2})f^{t_{2'},t_{1'}|t_2,t_1}_{NN}(p'_{2\perp}-p_{2\perp})\over 
\Delta^0 + p_{mz}-p_{1z}+i\varepsilon} \nonumber \\
& \times & j_{t1}^{\mu}(p_1+q,s_f;p_1,s_1) 
\cdot \Psi^{s_A}_{A}(p_{1},s_1,t_1;p_{2},s_2,t_2;p_{3},s_3,t_3),
\label{FSI2c}
\end{eqnarray}
where $\chi_1(s^{NN}_{a2}) = {\sqrt{s^{NN}_{a2}(s^{NN}_{a2}-4m^2)}\over 2qm}$ 
with $s^{NN}_{a2} = (p_1+q+p_2)^2$ and
$\chi_2(s^{NN}_{b3}) = {\sqrt{s^{NN}_{b3}(s^{NN}_{b3}-4m^2)}\over 2p_{fz}m}$ with 
$s^{NN}_{b3} = (p_1+q+p_2-p'_{2}+p_3)^2$.

To complete the calculation of the double rescattering amplitude one 
should calculate also the amplitude corresponding to the diagram of Fig.{\ref{FSI2fig}(b).
This amplitude is obtained by interchanging momenta  of nucleons ``2'' and ``3'' in Eq.(\ref{FSI2c}). 
Furthermore, it is more convenient to express the integrand of the double rescattering 
amplitude through the momentum transfers in the NN rescattering amplitude
$k_2 = p'_2-p_2$ and $k_3 = p'_3-p_3$. Using these variables, and  changing the  $d^3p'_3$  
integration to $d^3p'_{23}$ (similar to what was done in Sec.\ref{IAsec}), one obtains for 
the complete  double rescattering amplitude:
\begin{eqnarray}
& &  A_{2}^{\mu}   =   A_{2a}^{\mu} +A_{2b}^{\mu}   = \nonumber \\
& & \ \ \ \ {F \over 4}\sum\limits_{s_{1},s_{2},s_{3}}\sum\limits_{t_1,t_2,t_3,t_{1'},t_{2'},t_{3'}}
\int d^3 p'_{23}  {d^3k_3\over (2\pi)^3}{d^3 k_2 \over (2 \pi)^3} 
\Psi_{NN}^{\dagger p_{r23},s_{r2},t_{r2};s_{r3},t_{r3}}(p'_{23},s_{2},t_{2'};s_{3},t_{3'})  
\nonumber \\ 
& &\times \left[ 
{\chi_2(s^{NN}_{b3})f^{t_{3'},t_f | t_3, t_{1'}}_{NN}(k_{3\perp})\over \Delta_3 - k_{3z} + 
i\varepsilon} 
{\chi_1(s^{NN}_{a2})f^{t_{2'},t_{1'}|t_2,t_1}_{NN}(k_{2\perp})\over \Delta^0 -k_{2z}-k_{3z}+i\varepsilon}
+\right. \nonumber \\
& & \left. 
{\chi_2(s^{NN}_{b2})f^{t_{2'},t_{f}|t_2,t_{1'}}_{NN}(k_{2\perp})\over \Delta_2 - k_{2z} + i\varepsilon} 
{\chi_1(s^{NN}_{a3})f^{t_{3'},t_{1'} | t_3, t_{1}}_{NN}(k_{3\perp})\over \Delta^0 -k_{2z}-k_{3z}+i\varepsilon} \right]
j_{t1}^{\mu}(p_m+k_2+k_3+q,s_f;p_m+k_2+k_3,s_1) \nonumber \\
& & \times \Psi^{s_A}_{A}(p_{m}+k_3+k_2,s_1,t_1;-{p_m\over 2}-k_2+p'_{23},s_2,t_2;
-{p_m\over 2}-k_3 - p'_{23},s_3,t_3).
\label{FSI2fin}
\end{eqnarray}

\subsection{Differential Cross Section}

The calculated amplitudes in Sec.\ref{III} allow us  to evaluate numerous
observables (both polarized and unpolarized) for  the high $Q^2$ quasi-elastic 
electro-production  from $^3He$ target.
The differential cross section of reaction (\ref{reaction}) is given by
\begin{eqnarray}
d^{12}\sigma&=&\frac{1}{4j_A}(2\pi)^{4}\delta^{4}(k_{e}+P_{A}-k_{e}
^{\prime}-p_{f}-p_{r2}-p_{r3})\sum\limits_{nucleons}|M_{fi}|^2 \nonumber\\
&&\frac{d^3 k_{e}^{\prime}}{(2 \pi)^3 2E_{e}^{\prime}} 
\frac{d^3 p_{f}}{(2 \pi)^3 2E_{f}} \frac{d^3 p_{r2}}{(2 \pi)^3 2E_{r2}} 
\frac{d^3 p_{r3}}{(2 \pi)^3 2E_{r3}} \ ,
\label{cs1}
\end{eqnarray}
where $j_A = \sqrt{(k_eP_A)^2-m_e^2M_A^2}$. 
Here we sum over the nucleons knocked-out by virtual photon.

final  and average over initial spins. 
The factor 1/4 comes from the averaging over the initial polarizations of 
the electron and $^3He$. 
Since one of the recoil nucleons is not observed, one eliminates this degree 
of freedom by integrating over $d^3 p_{r3}$. Thus integrated differential 
cross section is
\begin{eqnarray} 
d^{9}\sigma&=&\frac{1}{4j_A}(2\pi)^{4}
\delta(E_{e}+M_{A}-E_{e}^{\prime}-E_{f}-E_{r2}-E_{r3})
\sum\limits_{nucleons}|M_{fi}|^2 
\nonumber\\
&&\frac{d^3 k_{e}^{\prime}}{(2 \pi)^3 2E_{e}^{\prime}} 
\frac{d^3 p_{f}}{(2 \pi)^3 2E_{f}} \frac{d^3 p_{r2}}{(2 \pi)^3 2E_{r2}} 
\frac{1}{(2 \pi)^3 2E_{r3}} \ ,
\label{cs2}
\end{eqnarray}
where
${\bf p_{r3}}={\bf k_{e}}-{\bf k_{e}^{\prime}}-{\bf p_{f}}-{\bf p_{r2}}$.
In Eqs.\ (\ref{cs1}) and (\ref{cs2}) the transition matrix, $M_{fi}$, represents the 
convolution of the electron and nuclear currents, in which 
the nuclear current represents the sum of the IA, single and double rescattering amplitudes,
\begin{equation}
M_{fi} = -4\pi\alpha {1\over q^2}j^e_{\mu}
\cdot\left(A_{0}^{\mu} + A_{1}^{\mu} + A_{2}^{\mu}\right),
\label{m}
\end{equation} 
where $A_{0}$, $A_{1}$ and $A_{2}$ are defined in Eqs.\ (\ref{IAfin}), 
(\ref{FSI1fin}) and (\ref{FSI2fin}) respectively.

\section{Summary}
\label{IV}
We develop a theoretical framework for calculation of high $Q^2$ exclusive electro-disintegration 
of $A=3$ system. The main feature of our approach is the calculation  of final state interactions 
of the struck energetic nucleon with the recoil nucleons within  the generalized eikonal 
approximation (GEA) which allows us to account for the finite and relatively large  momenta 
of the  recoil nucleons.
An important advantage of this approach is that we can now self-consistently study short range 
correlations in nuclei since the GEA does not require the stationary condition for recoil nucleons 
as it does  the conventional Glauber approximation.

To describe the residual interaction between two recoil nucleons, we use a  scattering 
representation of the two-nucleon continuum state wave function. This allows us to evaluate 
the latter through the NN scattering amplitudes in the low-to-intermediate energy region.

In the second part of our work\cite{SASF2} we discuss the numerical calculations based on the 
formulae derived in this  paper. In numerical calculations as an input we use the 
calculation of ground state wave functions of $^3He$ based on realistic $NN$ interaction potentials 
as well as including models that account for the  three nucleon forces\cite{Bochum}. 
The numerical estimates of 
the interaction between recoil nucleons - referred to as pair distortion- is implemented 
through the parameterizations of low-to-medium energy $NN$ scattering amplitudes provided 
by the SAID group\cite{SAID}. For high energy small angle $NN$ scattering we use the 
parameterization of the form of Eq.(\ref{fnn}). In numerical calculations\cite{SASF2} we 
are interested mainly in studies of short-range two- and  three- nucleon correlations for 
which GEA provides an appropriate theoretical framework.

\begin{acknowledgements}
We thank Ted Rogers for careful reading of the manuscript and for many valuable comments.
This work is supported  by  DOE grants under contract DE-FG02-01ER-41172 and DE-FG02-93ER40771 as 
well as by the Israel-USA Binational Science Foundation Grant.
M.M.S. gratefully acknowledges also a contract from Jefferson Lab under which this work was 
done. The Thomas Jefferson National Accelerator Facility (Jefferson Lab) is operated 
by the Southeastern Universities Research Association (SURA) under DOE contract 
DE-AC05-84ER40150.
\end{acknowledgements}

\appendix 
 
\vspace{0.4cm}
 
\section{Feynman Diagram Rules for the Scattering Amplitude in GEA}

Within the GEA the general $eA$ scattering amplitude of Fig.\ref{FDR} can 
be calculated based on effective Feynman diagram rules   
formulated as follows\cite{ggraphs,treview}:

\begin{itemize}
\item We assigns the vertex functions $\Gamma_{A}(p_1,...,p_A)$ and 
$\Gamma^\dag_{A-1}(p'_2,...,p_A)$ to describe the  transitions between 
$''nucleus~ A''$ to $''A~nucleons''$ with momenta $\{p_n\}$, 
$\{p'_n\}$ and $''(A-1)~nucleons''$ with momenta $\{p'_n\}$ 
to $``(A-1)~ nucleon~final ~state''$ respectively.

\begin{figure}[ht]
\centerline{\epsfig{height=4cm,width=9cm,file=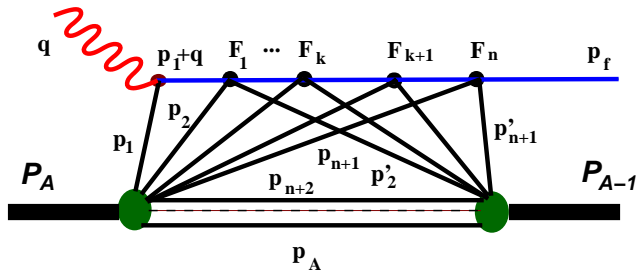}}
\caption{(Color online) n-fold $A(e,e'N)A-1$ scattering diagram}
\label{FDR}
\end{figure}

\item For $\gamma^* N$ interaction we assign vertex, $\Gamma^{h}_{\gamma^*N}$.
\item For each $NN$ interaction we assign the vertex function 
$F^{NN}_k(p_{k+1},p'_{k+1})$. This vertex function are related to 
the amplitude of $NN$ scattering as follows:
\begin{equation}
\bar u(p_3)\bar u(p_4)F^{NN}u(p_1)u(p_2) = 
\sqrt{s(s-4m^2)} f^{NN}(p_{3},p_{1})
\label{NN}
\end{equation}
where $s$ is the total invariant energy of two interacting nucleons with momenta 
$p_1$ and $p_2$ and 
\begin{equation}
f^{NN} = \sigma_{tot}^{NN}(i+\alpha^{NN})e^{-{B^{NN}\over 2}(p_3-p_1)_\perp^2},
\label{fnn}
\end{equation}
where $\sigma_{tot}^{NN}$, $\alpha$ and $B$ are known experimentally from $NN$ scattering 
data. The vertex functions are accompanied with $\delta$-function of energy-momentum 
conservation.

\item For each intermediate nucleon with four momentum $p$ we assign propagator 
$D(p)^{-1}=-(\hat p-m+i\epsilon)^{-1}$. Following to   
Ref.\cite{Gribov}  we choose the ``minus'' sign for the nucleon 
propagators to simplify the calculation of the overall sign of 
the scattering  amplitude.

\item The factor $n!(A-n-1)!$ accounts for  the combinatorics  
of $n$- rescatterings and $(A-n-1)$ spectator nucleons.

\item For each closed contour one gets 
the factor  ${1\over i(2\pi)^4}$   with no additional sign. 

\end{itemize}

Using above defined rules for the  scattering amplitude of Fig.\ref{FDR} 
one obtains:
\begin{eqnarray}
A^{(n)}_{A,A-1}(q,p_f) & = & \sum\limits_{h}{1 \over n!(A-n-1)!}
\prod\limits_{i=1}^{A} \prod\limits_{j=2}^{A} \int d^4p_i d^4p'_j 
{1\over \left[ i (2\pi)^4\right]^{A-2+n}}
\nonumber \\ 
& & 
\delta^4(\sum\limits_{i=1}^{A}p_i-{\cal P}_A) 
\delta^4(\sum\limits_{j=2}^{A}p'_j-{\cal P}_{A-1}) 
\prod\limits_{m=n+2}^{A}\delta^4(p_m-p'_m) \times 
\nonumber \\ 
& &  
{\bar u(p_f)\chi^\dag_{A-1}\Gamma^\dag_{A-1}(p'_2,..,p'_{n+1},p_{n+2},..,p_A)\over 
D(p'_2)..D(p'_{n+1})} 
{f^{NN}_n(p_{n+2},p'_{n+2}) .. f^{NN}_1(p_2,p'_2)
\over D(l_1)..D(l_k).. D(l_{n-1})}
\nonumber \\ 
& &  
{\Gamma^{h}_{\gamma^*N}(Q^2)\over  D(p_1+q)} 
{\Gamma_{A}(p_1,...,p_A)\chi_A
\over D(p_1)D(p_2)..D(p_{n+1})D(p_{n+2})..D(p_A)} 
\label{ampl_n} 
\end{eqnarray}
where, for the sake of simplicity, we  neglect the spin dependent  
indexes. Here,  ${\cal P}_A$ and  ${\cal P}_{A-1}$ are the four  momenta 
of the target nucleus and  final $(A-1)$ system, $p_i$ and $p'_i$  
are nucleon momenta in the nucleus $A$ and residual $(A-1)$ system 
respectively and $l_k=q+p_1+\sum\limits_{i=2}^{k}(p_i-p'_i)$.
The intermediate spectator states in the diagram of Fig.\ref{FDR}
are expressed in terms of nucleons but not nuclear fragments
because, in the high energy limit, the closure over  various nuclear excitations in the 
intermediate state is used\cite{ggraphs,treview}.
$\sum\limits_{h}$ in Eq.(\ref{ampl_n}) goes over virtual
photon  interactions  with different nucleons, in which  $\Gamma^{h}_{\gamma^*N}(Q^2)$ 
describes the  electromagnetic interaction. 

The vertex function $\Gamma_A$ describes a transition of nucleus-$A$ to 
the $A$-nucleon state, while $\Gamma^\dagger_{A-1}$ transition of the $A-1$ intermediate
nucleon to a final continuum or bound $A-1$ nucleon state.

If one considers the kinematic conditions  (similar to Eq.(\ref{kin})) in 
which the internal momenta of nucleons are restricted and the only relevant degrees 
of freedom are nucleons, one can evaluate  the intermediate state nucleon propagators
through the poles corresponding to the positive energy solutions. As a result
the  covariant amplitude  will be reduced to  a set of time ordered   
non covariant diagrams that allows us to establish the correspondence
between the nuclear vertex functions and the nuclear wave functions. 
In this limit the momentum space wave function is defined  through the vertex function as 
follows\cite{Gribov,Bert}: 
\begin{equation}
\psi_{A}(p_1,p_2,...,p_A)   =   {1\over \left(\sqrt{(2\pi)^3 2m}\right)^{A-1}}
{\bar u(p_1)\bar u(p_2)...\bar u(p_A)\Gamma_A(p_1,p_2,...,p_A)\over p_1^2-m^2}\chi^A,
\label{wf} 
\end{equation}
normalized as:   
\begin{equation}
\int|\psi_A(p_1,p_2,...,p_A)|^2 
\delta^3(\sum p_i-p_{\cal A})\prod\limits_{i=1}^{A} d^3p_i = N
\label{norm}
\end{equation} 
where $N=A$ for bound states and $N=\prod\limits_{i=1}^{A}\delta^3(p_i-p_i')$ 
for $A$ body continuum sate. Note that to apply the relativistic 
normalization for the spinors ($\bar u u = 2m$) the $\sqrt{(2\pi)^3 2 m}^{-1}$
factor should be associated with the plane wave single nucleon 
wave function.

 \end{document}